\documentclass[twocolumn,showpacs,preprintnumbers,amsmath,amssymb,showkeys]{revtex4}

\usepackage{graphicx}
\usepackage[]{subfigure}

\newcommand{\be}{\begin{equation}}
\newcommand{\ee}{\end{equation}}
\newcommand{\bea}{\begin{eqnarray}}
\newcommand{\eea}{\end{eqnarray}}
\newcommand  {\gtsim }  {\,\vcenter{\hbox{$\buildrel\textstyle>\over\sim$}}\,}

\newcommand{\fstage}{f_{\rm drive}}
\newcommand{\fsample}{f_{\rm sample}}
\newcommand{\fc}{f_{\rm c}}
\newcommand{\fco}{f_{\rm c,0}}
\newcommand{\fnu}{f_{\nu}}
\newcommand{\fm}{f_m}
\newcommand{\fmo}{f_{m,0}}

\newcommand{\fNyq}{f_{\rm Nyq}}

\newcommand{\DV}{D^{\rm volt}}

\newcommand{\Do}{D_0}

\newcommand{\kac}{\kappa_{\rm ex}}

\newcommand{\go}{\gamma_0}
\newcommand{\gac}{\gamma_{\rm ex}}
\newcommand{\gacbar}{\overline{\gamma_{\rm ex}}}
\newcommand{\gstokes}{\gamma_{\rm Stokes}}
\newcommand{\gfaxen}{\gamma_{\mbox{\scriptsize  Fax\'en}}}
\newcommand{\gfl}{\gamma}

\newcommand{\Tmsr}{t_{\rm msr}}
\newcommand{\kB}{k_{\rm B}}
\newcommand{\Ftherm}{F_T}
\newcommand{\Wth}{W_{\rm th}}
\newcommand{\Wthhydro}{W_{\rm th, hydro}}
\newcommand{\Wex}{W_{\rm ex}}

\newcommand{\Ptherm}{P_T}
\newcommand{\Pdriven}{P_{\rm response}}
\newcommand{\PV}{P^{\rm volt}}
\newcommand{\PVtherm}{P^{\rm volt}_T}
\newcommand{\PVdriven}{P^{\rm volt}_{\rm response}}

\newcommand{\Phydro}{P^{\rm hydro}}
\newcommand{\Pthermhydro}{P_T^{\rm hydro}}
\newcommand{\Pdrivenhydro}{P_{\rm response}^{\rm hydro}}

\newcommand{\xV}{x^{\rm volt}}
\newcommand{\xstage}{x_{\rm drive}}
\newcommand{\xdriven}{x_{\rm response}}
\newcommand{\xtherm}{x_T}
\newcommand{\vstage}{v_{\rm drive}}
\newcommand{\tphase}{t_{\rm lag}}
\newcommand{\tliquid}{t_{\rm phase}}

\newcommand{\degC}{$^{\circ}$C}
\newcommand{\decay}{\zeta}
\newcommand{\Aliquid}{A_{\rm liquid}}

\newcommand{\real}{{\rm Re}}
\newcommand{\imag}{{\rm Im}}

\begin{document}
%\preprint{PREPRINT}

%--- TITLE PAGE --------------------------------------------------------------------
\title{Calibration of optical tweezers with positional detection in the back-focal-plane}
\date{\today}

\author{Simon F. Toli\'{c}-N\o rrelykke\footnote[1]{corresponding author}\footnote[2]{These authors contributed equally to this work}}
\affiliation{Max Planck Institute for the Physics of Complex Systems,
N\"othnitzer Strasse 38, 01187 Dresden, Germany\\
and\\
European Laboratory for Non-linear Spectroscopy, via Nello Carrara 1,
50019 Sesto Fiorentino (Fl), Italy}

\author{Erik Sch\"{a}ffer\footnotemark[2]}
\affiliation{Max Planck Institute of Molecular Cell Biology and Genetics,
Pfotenhauerstrasse 108, 01307 Dresden, Germany}

\author{Francesco S. Pavone}
\affiliation{European Laboratory for Non-linear Spectroscopy, via Nello Carrara 1,
50019 Sesto Fiorentino (Fl), Italy}

\author{Jonathon Howard}
\affiliation{Max Planck Institute of Molecular Cell Biology and Genetics,
Pfotenhauerstrasse 108, 01307 Dresden, Germany}

\author{Frank J\"ulicher}
\affiliation{Max Planck Institute for the Physics of Complex Systems,
N\"othnitzer Strasse 38, 01187 Dresden, Germany}

\author{Henrik Flyvbjerg}
\affiliation{Isaac Newton Institute for Mathematical Sciences, Cambridge, U.K.\\
and\\
Biosystems Department and Danish Polymer Centre, Ris{\o} National Laboratory,
DK-4000 Roskilde, Denmark}
%------------------------------------------------------------------------------------------

%--- ABSTRACT------------------------------
\begin{abstract}
We explain and demonstrate a new method of force- and
position-calibration for optical tweezers with back-focal-plane
photo detection. The method combines power spectral measurements of
thermal motion and the response to a sinusoidal motion of a
translation stage. It consequently does not use the drag coefficient
of the trapped object as an input. Thus, neither the viscosity, nor
the size of the trapped object, nor its distance to nearby surfaces
need to be known. The method requires only a low level of
instrumentation and can be applied \textit{in situ} in all spatial
dimensions. It is both accurate and precise: true values are
returned, with small error-bars. We tested this experimentally, near
and far from surfaces. Both position- and force-calibration were
accurate to within 3\%. To calibrate, we moved the sample with a
piezo-electric translation stage, but the laser beam could be moved
instead, e.g.\ by acousto-optic deflectors. Near surfaces, this
precision requires
an improved formula
for the hydrodynamical
interaction between an infinite plane and a micro-sphere in
\emph{non-constant} motion parallel to it. We give such a formula.
\end{abstract}
%\pacs{87.80.-y, 06.20.Dk, 07.60.-j, 05.40.Jc}

\keywords{Optical tweezers, calibration, piezo-stage, acousto-optic deflectors,
back-focal-plane, imaging detectors, hydrodynamic interaction}

\pagestyle{myheadings}
\markright{Calibration of optical tweezers}

\maketitle

%%%%%%%%%%%%%%%%%%%%%%%%%%%%%%%%%%%%%%%%%%%%%%%
\section{INTRODUCTION}

In order to use optical tweezers as a quantitative instrument for position and force measurements,
the detection system must be calibrated.
One calibration method interprets the power spectrum of thermal Brownian motion
of a trapped object~\cite{Neuman2004,Berg-Sorensen2004,Berg-Sorensen2006_preprint}.
Another calibration method interprets the displacement of a trapped object
in response to a known flow past it~\cite{Kuo1993,Simmons1996}.
Both these methods require that the drag coefficient of the trapped object is known.
Here we combine the two methods into one.
Then there is no need to know the drag coefficient.

Mathematically speaking, the combined method measures an extra quantity
that allows the elimination of the drag coefficient from the calibration procedure.
This has some advantage:
Methods that use the drag coefficient as input,
use Stoke's law to calculate it,
hence rely on assumptions about the object's shape and radius,
and about the viscosity of the surrounding fluid.
This contributes to the error on the final calibration,
as do hydrodynamical interactions with nearby surfaces \cite{footnote1}.

With the combined method presented here, these sources of error have been eliminated.  The drag-coefficient of the trapped object can in fact be measured with the present method, and this with precision, as shown below.
Finally, the method presented here is simpler to implement and requires less instrumentation than other methods that use additional lasers \cite{Smith1996,Lang2002} or AODs \cite{Vermeulen2006}.

%%%%%%%%%%%%%%%%%%%%%%%%%%%%%%%%%%%%%%%%%%%%%%
\section{MATERIALS AND METHODS}
\label{sec:materials} Measurements were done with two different
optical tweezers systems. Both use the trapping laser for position
detection in the back focal plane. One system has a long working
distance and was used to test the method far from surfaces
(30\,$\mu$m): This experimental setup is described in detail
in~\cite{Capitanio2005} and briefly here. The instrument is  based
on a custom-built inverted microscope with a Nikon, 60$\times$, 1.2
NA, 0.2\,mm working distance, Plan-Apo, water immersion objective.
The laser is a 1064\,nm, Nd:YAG (Spectra-Physics Millennia IR\@).
Position detection is done with a position-sensitive photodiode (UDT
DLS-20). Flow-cells with a volume of 10\,$\mu$l (dimensions $8
\times 20 \times 0.06$\,mm$^3$) were assembled by placing a
coverslip on top of a microscope slide separated by spacers
of double-sided sticky tape. A dilute solution of beads was flowed in,
and the ends were sealed with nail polish to avoid sample
evaporation. The flow-cell was mounted upside down on a Physik
Instrumente piezo-electric translation stage (P-527.2 C1).

The other system has a short working distance and was used to test the method close to surfaces (0--3\,$\mu$m):
This experimental setup is described in detail in \cite{Schaeffer_in_prep}.
Briefly, it consists of a modified Zeiss Axiovert 135 TV microscope
equipped with a Zeiss, 100$\times$, 1.3 NA, Plan-Neofluar, oil-immersion objective.
The laser is a 1064\,nm, Nd:YVO$_4$ (Smart Laser Systems GmbH, Berlin, Germany).
Position detection is obtained with a standard quadrant photodiode, QP50-6SD (Pacific Silicon Sensors Inc).
Signals were recorded with a 24-bit data acquisition card (NI 4472, National Instruments)
which has a 45\,kHz alias-free bandwidth.
Temperature was measured with type-T thermocouples, accurate to within 0.1\,\degC \ (IT-23 Physitemp, Clifton, NJ, USA).
Flow-cells with a 3\,mm-wide channel were assembled by placing a 18\,mm$^2$ cover-slip on top of a No.\,1.5 22\,mm$^2$ cover-slip separated by a layer of parafilm.
The parafilm was melted by placing the sample on a 100$^{\circ}$C hot-plate.
Cooling then glued the cover-slips together.
A dilute solution of microspheres was flowed in,
and the ends were sealed with vacuum grease to avoid sample evaporation.
The flow-cell was mounted on a Physik Instrumente piezoelectric translation stage (P-733.2 CL) with built-in capacitance position detection pre-calibrated to within 0.1\%.
For measurements close to the surface,
the cover-slips were treated as described in \cite{Schaeffer_in_prep} to reduce the influence of surface potentials.

The following beads were used:
Silica beads from Bangs Laboratories (9025 Technology Drive, Fishers, IN 46038-2886, USA), catalog code SS04N, lot number 5303,
were 1.54\,$\mu$m  in diameter with a coefficient of variation (standard deviation divided by the mean value) of 10\% listed by the manufacturer.
Polystyrene microspheres from Polysciences  (Warrington, PA 18976, USA), catalogue number 07307, lot number 50602,
were 528\,nm in diameter with a 2\% coefficient of variation listed the manufacturer.
Transmission electron microscopy (TEM) of the latter microspheres indicated a 1.2\%\ coefficient of variation.

%%%%%%%%%%%%%%%%%%%%%%%%%%%%%%%%%%%%%%%%%%%%%%%%%%%%%%%%%%%%%%%%%%%%%%%%%%%%%%%%%%%%%%%%%
\section{Theory}
\label{sec:theory}
In the following we present the theory for a calibration procedure
in which the flow-cell is moved sinusoidally a known distance relative to the trapping laser
by a piezo-electric translation stage.
If, instead, the trap is moved a known distance relative to the flow-cell,
the same formulae apply,
provided the detector is placed in the back focal plane of the condenser.

In order to keep the presentation as simple as possible,
we use Einstein's simple theory for Brownian motion.
The presentation given here carries over unchanged to the full, hydrodynamically correct theory,
including possible filters, electronic and/or parasitic \cite{Berg-Sorensen2004,Berg-Sorensen2006_preprint}.
We present the relevant formulas in Appendix~\ref{app:Phydro}.

\subsection{Equation of motion}
\begin{figure}[ht]
    \includegraphics[width=0.9\linewidth]{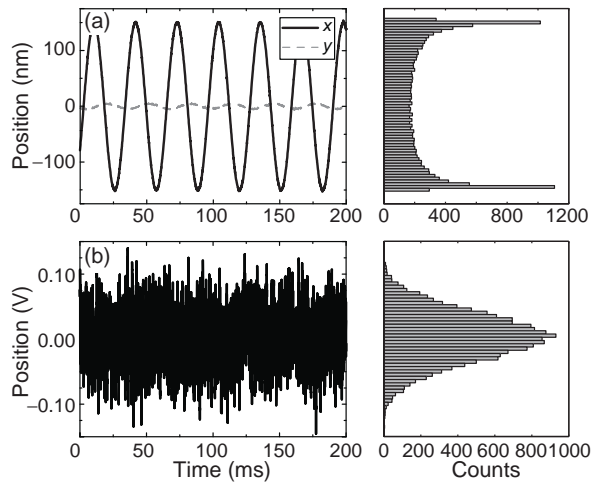}
    \caption{
    Positions of the piezo-stage (a) and the trapped bead (b)
   for a stage moving sinusoidally with frequency $\fstage = 32$\,Hz and amplitude $A=150$\,nm in the $x$-direction.
(a): Left,  time series of stage position.  Right,  histogram of  $x$-coordinate of the stage position.
The sinusoidal movement results in two clear maxima.
    (b): Left, time series of the bead's $x$-coordinate in volts, as given by the signal from the photodiode.
    The amplitude of the sinusoidal response is smaller than the amplitude of the thermal motion.
    Consequently the sinusoidal shape is masked by the Brownian motion of the bead,
    and the maxima associated with the sinusoid disappear in the histogram of visited positions (right).
  } \label{fig:drive_signal}
   \end{figure}

A microsphere suspended in water is trapped with optical tweezers inside a flow-cell.
The stage moves the flow-cell sinusoidally relative to the optical trap
with a frequency $\fstage$ and an amplitude $A$, see Fig.~\ref{fig:drive_signal},
while the trap remains at rest in the laboratory system.
The position of the stage as a function of time $t$ is
\be
    \xstage(t) = A \sin( 2 \pi \fstage t )
    \enspace.
\ee
%---------------------------------------------------------------------------------------------------------
%---------------------------------------------------------------------------------------------------------
The stage velocity $\vstage(t) \equiv \dot{x}_{\rm drive}(t)$
also corresponds to the velocity of the water in the flow-cell far away from the bead,
since the water is at rest relative to the flow-cell (see Appendix~\ref{app:hydro}).
Ignoring hydrodynamical and inertial effects, the Langevin equation of motion for a spherical bead in the trap is
\be \label{eq:newtonlangevin}
    \gamma \left[ \dot{x}(t) - \vstage(t) \right] + \kappa x(t) = \Ftherm(t)
    \enspace
\ee
where $x(t)$ is the position of the bead relative to the center of the trap,
$\gamma$ is the drag coefficient, and $\kappa$ is the trap stiffness.
The first term on the left-hand-side is the drag force,
which is proportional to the velocity of the bead relative to that of the stage.
The second term on the left-hand-side is the trapping force.
The right-hand side is the random thermal force driving the Brownian motion.
It is assumed to have the statistical properties of white noise,
\be
    \Ftherm(t) = \sqrt{2\gamma \kB T} \, \xi(t) = \gamma \sqrt{2D} \, \xi(t)
\ee
where $\kB T$ is the Boltzmann energy at absolute temperature $T$,
the diffusion coefficient is given by Einstein's relation $D=\kB T/\gamma$,
and $\xi$ is the normalized white noise expressed with Dirac's delta function $\delta$,
\be
    \langle \xi(t) \rangle = 0\, ;  \;\;   \langle \xi(t) \xi(t') \rangle = \delta(t-t')
\enspace.
\ee
Here $\langle \ldots \rangle$ denotes the expectation value that results from averaging over the thermal noise.

\subsection{Solution to the equation of motion}
Since the equation of motion (\ref{eq:newtonlangevin}) is linear with two force terms,
 $\Ftherm$ and $\gamma  \vstage$, its general solution can be written as a sum of two terms, one for each force,
\be
    x(t) =   \xtherm(t) + \xdriven(t)
    \label{eq:bead_pos}
    \enspace ,
\ee
after transient initial behavior has died out.
Here,
\bea
    \xtherm(t) &=&  \sqrt{2D} \int_{-\infty}^t \! \mbox{d}t'  e^{-2 \pi \fc (t-t')}\, \xi(t')
    \label{eq:xtherm}\enspace,\\
    \nonumber \\
    \xdriven(t) &=& \frac{ \xstage ( t - \tphase ) }{ \sqrt{ 1 + (\fc/\fstage)^2 } }
    \label{eq:xdriven}
    \enspace
\eea
where we have introduced the corner frequency $\fc=\kappa/(2\pi\gamma)$,
and  $\tphase = [\arctan(\fstage/\fc) - \pi/2] / (2 \pi \fstage)$.
Figure~\ref{fig:drive_signal}b shows an example of an experimentally determined trajectory $x(t)$ of a bead in a trap,
as described in Eq.~(\ref{eq:bead_pos}).
In this case, the stochastic thermal motion dominates and almost hides the driven, deterministic component of the motion.

In principle, we now could calibrate
by fitting $\xdriven(t)$ in Eq.~(\ref{eq:xdriven})
to data like those shown in Fig.~\ref{fig:drive_signal}b.
However, this is not a reliable procedure \cite{Bockelmann2002}.
Instead, we Fourier transform theory and data to the frequency domain
where parameters are determined with optimal precision
because the theory is simpler there.

\subsection{Power spectrum}
From Eqs.~(\ref{eq:bead_pos}--\ref{eq:xtherm}) it follows that the Fourier transform of $x(t)$ is
\bea
 \lefteqn{   \hat{x}(f) = \int_{-\infty}^{\infty}  \mbox{d}t \, e^{i 2 \pi f t} x(t)}&& \nonumber \\
    &=&  \frac{ \hat{\xi}(f) \sqrt{2D} } {2 \pi (f_c - if)}  \label{eq:xhat} \\
    &+&   \frac{ A     e^{i 2 \pi f \tphase} }{ 2i \sqrt{ 1 + (\fc/\fstage)^2 } }
    \left[ \delta(f+\fstage) - \delta(f-\fstage) \right] \nonumber
    \enspace ,
\eea
where $\hat{\xi}(f)$ is the Fourier transform of $\xi(t)$.
Consequently, the expectation value for the one-sided ($f\ge 0$)
power spectral density (PSD) of the bead positions
is \cite{footnote2}
\bea
    \label{eq:inf_power}
    P(f) &=&     \frac{ 2\langle |\hat{x}(f)|^2 \rangle }{ \Tmsr }
    =   \Ptherm(f) + \Pdriven(f) \nonumber\\
    \label{eq:Pf}\\
    &\underset{ \Tmsr \rightarrow \infty }{ \longrightarrow }&
    \frac{ D }{\pi^2 (f^2 + \fc^2)} + \frac{ \frac{1}{2}  A^2}
    { 1 + \fc^2/\fstage^2 }\delta(f-\fstage)   \nonumber
    \enspace
\eea
where $\Tmsr$ is the measurement time (see also \cite[Eq.~(8)]{Berg-Sorensen2004}).
This PSD consists of the familiar Lorentzian (first term, $\Ptherm$),
\emph{plus} a delta-function spike (second term, $\Pdriven$)
at the frequency with which the stage is driven.
The Lorentzian originates from the Brownian motion of the bead in the parabolic trapping potential,
and is hereafter referred to as the ``thermal background.''

%%%%%%%%%%%%%%%%%%%%%%%%%%%%%%%%%%%%%%%%%%%%%%%%%%%%%%%%%%%%%%%%%%%%%%%%%%%%%%%%%%%%%%%
\section{How to Calibrate}
\label{sec:finite}

Experimentally, positions are measured in volts, $\xV$.
Assuming linearity \cite{Schaeffer_in_prep,Capitanio2005},
\be \label{eq:linearity}
	x(t) = \beta \, \xV(t) \enspace,
\ee
positions are known in meters once the calibrations factor $\beta$ has been determined.
It can be determined from the measured PSD:
From Eq.~(\ref{eq:linearity}) follows that the experimental PSD $\PV$ is measured in (volts)$^2\times$(seconds) and
\be \label{eq:PbP}
	\Pdriven(f) = \beta^2 \PVdriven(f) \enspace.
\ee
Here, $\PVdriven$ is known experimentally and, as can be seen from Eq.~(\ref{eq:inf_power}), so is $\Pdriven$, since $A$ and $\fstage$ are known a priori, and $\fc$ is known experimentally.
So the calibration factor $\beta$ is the only unknown in Eq.~(\ref{eq:PbP}), hence determined by this equation.

In practice the experimental measurement time is always finite.
This is taken into account in the expression for  $\beta$ given below.

The procedure described here works in bulk and near a surface.
If used near a surface, the resulting calibration is specific to
the distance from the surface.
So is the value found for the drag coefficient.
Fax\'en's law (Eq.~(\ref{eq:faxen}) below) and formulas in \cite[Sect.~7.4]{HappelBrenner,Schaeffer_in_prep} relate the drag-coefficient to its bulk value for the lateral and axial direction, respectively.

\subsection{Positional calibration}
\begin{figure}[ht]
  \includegraphics[width=0.9\linewidth]{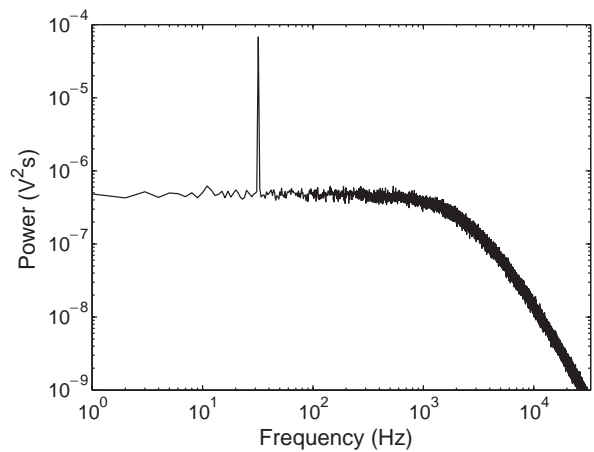}
    \caption{ Power spectrum of a 528\,nm diameter polystyrene bead held in
    the laser trap with a corner-frequency $\fc = 2065\pm5$\,Hz.
    The sample moves sinusoidally with amplitude $A=150$\,nm and
    frequency $\fstage = 32$\,Hz.
    The power spectrum shown is the average of 100 independent power spectra.
    It consists of a thermal background caused by Brownian motion,
    plus a spike at $\fstage$.
    The sampling frequency was $\fsample  = 65\,536$\,Hz, the measurement time for
    each spectrum $\Tmsr = 1$\,s, and the temperature 24.4$^{\circ}$\,C.
    For clear illustration, the measurement time used here is eight times longer
    than the one we typically use for calibration.}    \label{fig:bead_power_spectrum}
\end{figure}

The desired calibration factor for distances is
\be
    \beta = \sqrt{\Wth/\Wex} \, , \,\,\, [\beta] =  \mbox{m/V}
    \enspace,
    \label{eq:conversion}
\ee
where $\Wex$ is the experimentally determined power in the spike, measured in (volts)$^2$,
and $\Wth$ is the same quantity measured in (meters)$^2$.
The latter follows from  Eq.~(\ref{eq:inf_power}),
\be
\Wth = \int_0^{\fNyq} \!  \Pdriven(f)\,\mbox{d}f
     = \frac{ \frac{1}{2} A^2}{1+\fc^2/\fstage^2}
     \enspace ,
\label{eq:Wth}
\ee
whereas $\Wex$ is found by observing
that in an experiment the measurement time $\Tmsr$ is not infinite,
so the spike is no longer a delta-function.
As detailed in Appendix~\ref{app:psd_rect},
its height is now proportional to $\Tmsr$,
and its width is also finite in general.
However, if
$\Tmsr$ is an \emph{integer multiple} of the period of the stage movement,
then the spike consists of a single datum \cite{Harris1978}, see Fig.~\ref{fig:bead_power_spectrum}, and consequently
\be
    \label{eq:Wex}
    \Wex = \left( \PV(\fstage) -   \PVtherm(\fstage) \right) \Delta f
    \enspace,
\ee
where $\PV(\fstage)$ is the height of the spike,
i.e., the experimentally determined value of the PSD at $\fstage$ and
$\PVtherm(\fstage) $ is the PSD of the thermal background at $\fstage$.
Both are measured in (volts)$^2 \times$(seconds).
The width $\Delta f =  1/\Tmsr$ corresponds to the frequency resolution of the PSD.
%--------------------------------------------------------------------------
%--------------------------------------------------------------------------

The thermal background in this expression can be treated with three
different levels of precision:
(i)~Do not subtract it at all, if it
is truly negligible compared to the height of the spike.
(ii)~Interpolate its value from a plot of the power spectrum. This
approach can also be taken if the PSD is contaminated with
low-frequency power, that is external to the model used here, e.g.\
from electronic noise or the pointing instability of the laser beam.
(iii)~Use the value at $\fstage$ of the theoretical expression for
the thermal background $\PVtherm(f)$
after it has been fitted to the experimental thermal background.

%%%%%%%%%%%%%%%%%%%%%%%%%%%%%%%%%%%%%%%%%%%%

\subsection{Force calibration}
The trap's force on a bead is $\kappa x$.
With the position detection system calibrated, the displacement $x$ is measured in meters.
To determine the trap stiffness $\kappa$,
we use the definition of $\fc$ and Einstein's relation to write $\kappa = 2\pi\fc \kB T/ D$.
Next, we determine $\fc$ and $\DV$, the diffusion constant in (volts)$^2$/(seconds),
 by fitting the first term in Eq.~(\ref{eq:inf_power}) to the thermal background
 in the experimental PSD.
 Using $D=\beta^2 \DV$, we arrive at a practical formula for force-calibration

 \be   \label{eq:kac}
\kac  = 2\pi\fc  \frac{\kB T}{\beta^2 \DV}
\enspace.
 \ee
All variables on the right-hand side are measured experimentally
 in the calibration process, including $\Wth$, which is determined as the right-hand side of
 Eq.~(\ref{eq:Wth}).
The local temperature $T$ of the liquid can typically be determined with sufficient accuracy
by direct measurement in or near the flow-cell:
$T$ is the absolute temperature,
so an absolute error of 1\,K results in a relative error of only 0.3\%.

As a side effect of this calibration procedure, we find an
experimental result for the drag coefficient,
\be  \label{eq:gac}
\gac  =   \frac{\kB T}{\beta^2  \DV}
\enspace,
 \ee
where again all variables on the right-hand side are measured experimentally
 in the calibration process.

Use of the hydrodynamically correct theory for Brownian motion
changes Eq.~(\ref{eq:conversion}) only by replacing $\Wth$ with $\Wthhydro$ (see Appendix~\ref{app:Phydro}).
The values found for $\kac$ and $\gac$ with these equations do change, however,
because more correct values
for $\fc$ and $\DV$ result from fitting the thermal background
with the hydrodynamically correct theory.
The value found for $W_{\rm{ex}}$ is also affected a little
through the subtraction of the thermal background in Eq.~(\ref{eq:Wex}).

%%%%%%%%%%%%%%%%%%%%%%%%%%%%%%%%%%%%%%%%%%%%%%%
\section{Experimental results}
\label{sec:test}
In order to test the accuracy and advantage of the calibration method described above,
we here compare our experimentally determined values for the drag coefficient
with values calculated from vendor information.
Deep in bulk, 21 silica beads with a diameter of 1.54\,$\mu$m were studied
with an optical tweezers system with a water-immersion objective.
At various distances close to the cover slip ($< 3$\,$\mu$m),
24 polystyrene beads with a diameter of 528\,nm were studied
 with an optical tweezers system with an oil-immersion objective.
Fitting of the PSD to the thermal background recorded for each bead at each of its positions considered
 was done with the highest possible precision,
using either published MatLab routines \cite{Tolic-Norrelykke2004,Hansen2006}
or custom written software in LabView \cite{Schaeffer_in_prep}.
These fitting routines take into account hydrodynamics corrections,
aliasing, parasitic filtering in the photo diode,
and electronic filters in the data acquisition system~\cite{Berg-Sorensen2004,Berg-Sorensen2006_preprint}.

%%%%%%%%%%%%%%%%%%
\subsection{Measurements in bulk}
First we measured the drag coefficient far from surfaces,
where $\gac$ can be compared directly to Stokes's formula $\go = 6\pi\eta R$,
with the bead diameter $2R=1.54$\,$\mu$m taken from the specifications of the producer,
and the viscosity $\eta=0.93$\,mPa$\cdot$s calculated for water at the measured temperature, 23.0\,\degC.
Beads were trapped near the bottom of the flow-chamber (silica beads are heavier than water)
and brought to the middle of the flow-cell, 30\,$\mu$m from the bottom and the top,
to minimize the effect of nearby surfaces.
The $x$-axis---the direction of motion---was chosen perpendicular to the long axis
of the flow-cell and perpendicular to the direction of the incoming laser light (the $z$-axis).
The experimental parameters were $A=208$\,nm and $\fstage=28$\,Hz.
For each of the 21 beads, $\fc$, $\DV$, and $\Wex$ were determined from its power spectrum
recorded in bulk.

The average of values measured for $\gac$ was $\gacbar=(13.4\pm0.2)$\,nN$\cdot$s/m
(mean $\pm$ standard error on the mean (SE), $n=21$ beads),
and the SD of measured values for $\gac$ was 0.8\,nN$\cdot$s/m,
i.e., the measured drag coefficient showed a coefficient of variation of 6\%,
which agrees with the 10\% variation in bead diameter listed by the manufacturer
because of our limited sample size.
The  value for the average drag coefficient expected from the manufacturer's information is
$(13.5 \pm 0.3)$\,nN$\cdot$s/m (mean $\pm$ SE)
where the SE was obtained by adding the errors associated with viscosity and
variation in bead diameter in quadrature divided by $n$.
These results suggest that our calibration method is both accurate and precise.

%%%%%%%%%%%%%%%%%%%%%%
\subsection{Measurements near surfaces}  \label{subsect:nearsurfaces}

In practice, experiments are often done close to a surface, such as a cover slip.
Working close to a surface---in our case $<3\,\mu$m---complicates standard calibration techniques,
 especially if a high numerical aperture oil-immersion objective is used.
In that case aberrations arising from a refractive index mismatch at the glass-water interface
cause a linear decrease in stiffness away from the surface
and a focal shift \cite{Hell1993,Florin1998,Neuman2005}.

\subsubsection{Acquisition and fitting}
The experimental parameters were,
$A=150$\,nm, $\fstage=32$\,Hz, $\fsample=65\,536$\,Hz,  $\Tmsr=1/8$\,s,  and $T=24.4$\degC\@.
The beads used had, according to the producer,
a diameter of $2R=528$\,nm with a 2\% coefficient of variation.
Power spectra that resulted from averaging 100 independent spectra were fitted with a custom-written least-squares fitting routine
implementing a Levenberg-Marquardt algorithm (Labview, NI).
Each datum was weighted by its theoretical error bar~\cite{Berg-Sorensen2004}.
The propagated error on the fit parameters was calculated as the square root of the diagonal elements
of the covariance matrix multiplied by the reduced $\chi^2$-value.
The fit was done in the frequency interval [8\,Hz\,:\,25\,kHz],
omitting the single datum at the stage frequency $\fstage=32$\,Hz,
and using Eq.~(\ref{eq:Pthermhydro}) for the thermal background,
with Eq.~(35) in~\cite{Berg-Sorensen2004} describing parasitic filtering.

For every trapped bead the calibration factor $\beta$, the trap stiffness $\kac$,
and the drag coefficient $\gac$ were found at each of 50 distances from the surface.
The exact surface position was obtained from a fit of Fax\'en's law
to the measured drag coefficients, taking the focal shift into account;
see Appendix~\ref{app:Phydro} and \cite{Schaeffer_in_prep}.
Thus, the quoted values for $\gac$
are the measured values extrapolated to bulk,
hence directly comparable to $\go$.

\subsubsection{Temperature}
We measured the temperature with a small thermocouple introduced into the flow cell,
while simultaneously recording the room temperature and the temperature of
the imaging and condenser objective.
The temperatures of these two objectives differed by 0.5\,\degC\@.
The temperature inside the flow cell was intermediate between these and was measured to within 0.2\,\degC\@.
We estimated an upper limit of 0.5\,\degC\ for the local temperature increase due to laser heating \cite{Peterman2003}.
Thus, the propagated error on $\gac$ in Eq.~(\ref{eq:gac}) would be less than 0.2\% if uncertainty about the temperature
were the only source of error.

\subsubsection{Drag coefficient}
\begin{figure}[ht]
    \includegraphics[width=0.9\linewidth]{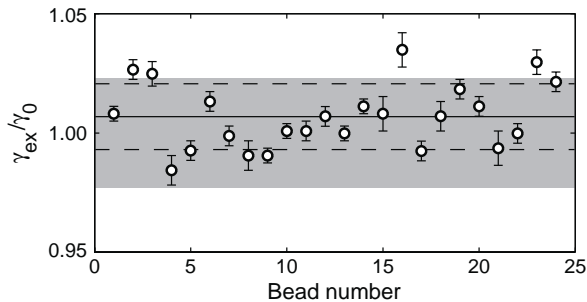}
     \caption{ \label{fig:bead_to_bead_variation}
     Results from the calibration of 24 individual beads trapped close to a surface.
    Circles show the measured drag coefficients Eq.~(\ref{eq:gac}),
    extrapolated to their bulk values according to Fax\'en's law,
    in units of the theoretically expected drag coefficient in bulk.
    Error-bars are the propagated errors from the fitting routines used.
   The horizontal solid line and the two dashed lines denote the mean ($\gacbar$) $\pm$ SD of the 24 measurements.
    The accuracy of the method is illustrated by the agreement of the two values $\gacbar/\go = 1.007 \pm 0.003$ (mean $\pm$ SE).
    The shaded region denotes the estimated 2.3\% uncertainty on $\go$.}
\end{figure}

Figure~\ref{fig:bead_to_bead_variation} shows the experimentally determined values for the drag coefficient,
$\gac$, here given in units of the theoretically expected drag coefficient, $\go$, for 24 individual beads.
The error-bars on the individual data-points are the propagated errors from the fit to Fax\'en's law
and the uncertainty in the temperature.
This high precision results from the long measurement time and the 50 determinations of $\gac$ for each bead.
Any systematic error---e.g.\ an undetected error in the specifications of the piezo-stage---will offset
the mean value of $\gac$, but will not change the position of the data-points relative to each other.
In other words, systematic errors directly influence the accuracy of the method, but do not influence the precision.
We check the accuracy by comparing $\gac$ to $\go$, including its estimated errors.
The shaded area shows the  2.3\% error on $\go$ from the propagated uncertainty on the viscosity (temperature) and the bead radius.
%---------------------------------------------------------------------------------
%---------------------------------------------------------------------------------

The average of the measured values
for the 24 beads, $\gacbar$,
was  $1.007 \pm 0.003$ (mean $\pm$ SE, $n=24$) in units of $\go$.
As seen in Fig.~\ref{fig:bead_to_bead_variation},
the experimental value for an individual bead may differ several percent
from the theoretically expected value $\go$, even if the average value does not.
Thus, if $\go$ is used for calibration,
stochastic errors of several percent are expected due to the poly-dispersity of the bead radii.

We were able to measure the poly-dispersity of the bead population
because the precision of our calibration method (error-bar on single-bead datum)
was smaller than this poly-dispersity.
The coefficient of variation of the bead-population's various $\gac$-values was 1.5\%,
which is comparable to the 2\% coefficient of variation for the bead diameter listed by the producer,
and the 1.2\% measured independently by us using TEM.

%%%%%%%%%%%%%%%%%%%%%%%%%%%%%%%%%%%%%%%%%%%%%%%%%%%%%%%%%%%%%%%%%%%%%%%%%%%%%%%%%%%%%%%%%
\section{Discussion}
\label{sec:discussion}

The main strength of the calibration method presented here is that it is independent of estimates
regarding the drag coefficient and that it can be performed on the location of the experiment.
It shares this strength with other methods~\cite{Lang2002,Vermeulen2006},
but unlike those it does not require two laser beams or acousto-optic deflectors.

The method presented here has some resemblance to a method used in \cite{Denk1990,Svoboda1993}.
There, a bead embedded in a gel was driven sinusoidally through the detection laser,
and the resulting spike in the PSD was used to estimate the sensitivity of the detection system.
However, that method was not \textit{in situ}
since the bead used for the calibration could not be used for later experiments.

In Appendix~\ref{app:issues} we discuss a number of issues associated with the use of optical tweezers in general
and our implementation of the calibration method in particular.
%%%%%%%%%%%%%%%%%%%%%%%%%%%%%%%%%%%%%%%%%%%%%%%%%%%%%%%%%%%%%%%%%%%%%%%%%%%%%
\section{Conclusions and Outlook}
\label{sec:conclusions}

\subsection{Recommended approach}
We suggest the following steps in experiments using a piezo-electric translation stage:
\begin{enumerate}
\item Trap the object of interest at the position of interest.

\item Drive the stage at
any  frequency $\fstage$ that does not excite resonances in the system.
We typically used 16\,Hz or 32\,Hz, low enough that the fluid moves with the stage, see Appendix~\ref{app:hydro}.
\label{stage}

\item Collect position data from the photo detection system and the piezo-stage
for a time $\Tmsr$ that is an integer number of stage periods---i.e., $\Tmsr \fstage =$ integer---to avoid leakage, and with a sampling frequency that makes the number of data points, $\Tmsr \fsample$, a power of two, so the PSDs of positions can be obtained by FFT.

\item Determine $A$ and $\fstage$ from the PSD of the piezo-stage positions (or use previously calibrated values).

\item Determine $\DV$ and $\fc$ from a fit to the thermal background in the PSD of the bead-positions\label{enum:fc}.

\item Determine the value of the spike in the PSD at $\fstage$ and calculate $\beta$ and $\kappa$
using Eqs.~(\ref{eq:conversion}) and (\ref{eq:kac}).
\end{enumerate}

In Point~\ref{stage},
the suggested values for $\fstage$ are powers of two for ease of calculation of the appropriate $\Tmsr$ and $\fsample$\@.
In Point~\ref{enum:fc}, we recommend using the full theory for fitting the PSD if less than 10\% systematic error is desired.
The relevant equations are given in Appendix~\ref{app:Phydro} and in \cite{Berg-Sorensen2004,Berg-Sorensen2006_preprint},
and software in \cite{Tolic-Norrelykke2004,Hansen2006}.

%%%%%%%%%%%%%%%%%%%%%%%%%%%%%%%%%%%%%%%%%%%%%%%%%%%%%%%%%%

\subsection{Calibration of the axial direction}
The total intensity of the laser light that impinges on the position
detector in the back-focal plane, contains information about the
displacement of the trapped object in the axial direction, i.e., in the
direction of the optical axis, the $z$-axis, \cite{Pralle1999}. It
is generally possible to calibrate this dimension by moving the
sample relative to the laser along the $z$-axis. If acousto-optic
deflectors are used to move the laser relative to the trapped object
\cite{Vermeulen2006}, this is impossible. However, for spherical
objects the bulk drag coefficient is the same in all directions, and
it is sufficient to determine the drag coefficient in one dimension
in order to calibrate all three dimensions. When working close to a
surface, the drag coefficient depends on the direction of motion.
However, if the bead is spherical and the distance to the surface is
known or determined as, e.g.\ in
Subsect.~\ref{subsect:nearsurfaces},
 it is again sufficient to determine the drag coefficient in just one direction.
 Fax\'en's law and formulas given in \cite[Sect.~7.4]{HappelBrenner,Schaeffer_in_prep} can be used to correct for the distance dependence
 of the drag coefficient for the remaining lateral and/or axial dimensions, respectively,
 depending on which dimension was previously calibrated.

\subsection{Imaging versus non-imaging positional detection}
Here we demonstrated our method using positional detection in the non-imaging back-focal-plane of the condenser.
In the back-focal plane an interference pattern is detected.
This pattern arises as the laser light scattered from the bead interferes with the unscattered laser light.

Position detection in the image plane should also work,
if the laser-trap remains stationary in the laboratory coordinate system,
while the flow cell is driven.
For example, an image of the bead could be projected onto a photodiode or a camera.
With such a setup, calibration can also be obtained by simply moving the detector a known distance relative to the image while recording the response \cite{Howard1988}.

Another approach  moves the trapping laser relative to the flow-cell,
e.g.\ using acousto-optic deflectors, galvano-mirrors,
or some other beam-steering apparatus \cite{Neuman2004}.
This approach works, if back-focal-plane detection is used:
A pure translation of the laser in the image-plane produces no signal in the back-focal-plane,
only a motion of the trapped object relative to the laser is detected.
An advantage of this approach is that open samples can be used, because the sample is not moving.

\subsection{\textit{In situ} measurements}
The method presented here is implemented strictly \textit{in situ}.
Therefore, it should be applicable in situations
that have so far eluded accurate measurements of positions
and forces, e.g.\ when trapping spherical structures
of unknown refractive index and size in the interior of cells.
The method described here could also be used in
microfluidic lab-on-a-chip devices, for measuring forces,
viscosities, or temperatures inside micron-sized channels.
Generally, position and force calibration should be possible in an arbitrary geometry,
because we do not need to know the corrections to the drag-coefficient due to the proximity of surfaces
in the approximation used in Eq.~(\ref{eq:newtonlangevin})
where the frequency-dependence of the drag-coefficient  is neglected.
More generally, all that is needed for the present method to apply with precision
is a system that displays a linear response to forces,
as in Eq.~(\ref{eq:newtonlangevin}):
The drag should be proportional to the trapped object's velocity,
but may depend on frequency as in Appendix~\ref{app:Phydro}
as long as the functional form of the frequency dependence is known,
or can be modelled, and the trapping force should be Hookean.

%%%%%%%%%%%%%%%%%%%%%%%%%%%%%%%%%%%%%%%%%%%%%%
\section{Acknowledgments}

HF and FJ thank the \textit{Isaac Newton Institute for Mathematical Sciences}
and its  program
\textit{Statistical Mechanics of Molecular and Cellular Biological Systems}
for hospitality while this work was initiated.
HF thanks Steve Block for the opportunity to present an early version of the present work
at his Winter Workshop on Biophysics, \textit{The Biophysics of Single Molecules}, Aspen, CO, January 2--8, 2005.
SFT-N thanks Marco Capitanio for technical assistance.
We all thank the anonymous reviewer for a thoroughly critical effort
that improved the presentation a good deal.
%%%%%%%%%%%%%%%%%%%%%%%%%%%%%%%%%%%%%%%%%%%%%%%
%\clearpage
\appendix
%%%%%%%%%%%%%%%%%%%%%%%%%%%%%%%%%%%%%%%%%%%%%%
\section{Experimental issues}
\label{app:issues}
We considered the following issues in the implementation of our calibration method.

\begin{enumerate}
\item \emph{Stage response.}
Piezo stages have a finite response time, but this is not a problem here,
since we drive with a sinusoidal signal.
This is the signal, which another periodic signal will degenerate towards,
if its period is shorter than the response time of the stage.
Also, any deviation from the chosen signal shows up in the experimental power spectrum as distinct,
isolated higher modes at frequencies that are integer multiples of the driving frequency.

Mechanical resonances of the experimental setup can be exited.
Our setups have resonance frequencies starting at approximately 400\,Hz.
The excitation of a resonance will give rise to additional power in the PSD at the resonance frequency, hence is under experimental control.
Mechanical crosstalk between axes occurs, as seen in Fig.~\ref{fig:drive_signal}.
However, as the equations of motion are linear, this does not influence the calibration.

\item \emph{Photodiode response.}
Photodiodes may act as filters,
but the effect is well understood and can be measured and accounted for~\cite{Berg-Sorensen2003,Berg-Sorensen2004,Berg-Sorensen2006_preprint}.
Depending on the type of diode used (quadrant or position sensitive), the linearity of the response to bead displacements may also vary.
 The region of linearity is easily found by moving a stuck bead through the laser focus.
The photodiode must be aligned with the piezo-stage.
This is achieved by driving the stage along one of its axes while rotating the diode until the power at the driving frequency is maximized for the corresponding axis.

\item \emph{Crosstalk.}
Crosstalk between the axial ($z$) and the lateral ($x$ and $y$)
channels from a quadrant-photodiode may lead to underestimates of $\fc$ of up to $10$\%.
This crosstalk shows up as additional power in the PSD of the $x,y$-channels.
Since the axial corner frequency is smaller than the lateral ones \cite{Rohrbach2005},
 crosstalk raises the plateau of the lateral signals.
A likely source of the crosstalk is differences in amplification of the signals from the diode's four quadrants.
By repositioning the diode relative to the laser
so that a small offset from the center position (i.e.\ zero volts in $x$ and $y$) is introduced,
the crosstalk can be minimized by maximizing the lateral corner frequencies
that are returned from fits to the PSD~\cite{Schaeffer_in_prep,Berg-Sorensen2004}.

\item \emph{Laser heating of the liquid.}
The laser heats up the liquid locally, resulting in a decrease in viscosity.
We looked for this effect by varying the laser intensity, but did not find any such effect when working close to surfaces.
This result is to be expected because the glass cover-slip acts as a heat-sink \cite{Peterman2003}.

\item \emph{Hydrodynamic response of the sample.}
The calibration method presented here is conditioned on the liquid co-moving with the stage.
In Appendix~\ref{app:hydro} we calculate the response of the liquid to the oscillatory movement of the stage.
Close to the surface, the no-slip boundary condition entrains the liquid.
Further into the sample, the degree of entrainment depends on the height of the sample $d$ and the drive frequency.
If $d^2 \pi \fstage < \nu \approx 1$\,mm$^2$/s,
where $\nu$ is the kinematic viscosity, the liquid co-moves in the entire flow-cell.

\item \emph{Shape of the trapped object.}
The only demand on the trapped object is that it does not rotate in response to forced movement,
at least not in a manner that gives rise to a response in the detection system.
This condition is fulfilled if the particle is spherical,
or asymmetric but strongly trapped.
Due to the height dependence of the drag coefficient,
even a spherical bead rotates when translated close to a surface.
However, this rotation dissipates a negligible amount of energy compared to the translational dissipation \cite{HappelBrenner} and hence does not affect the calibration.
If the trapped object is elongated and weakly trapped,
it may wobble in the trap in response to the oscillating liquid,
and give rise to detection of false movement by the photodiode.
But this is not a problem with commercial micro-spheres because they are highly spherical, as are many small biological objects, such as lipid droplets or vesicles.

\item \emph{Shape of the trapping potential.}
Throughout this paper we assumed a parabolic trapping potential, but the method is not limited by this assumption.
By choosing large drive amplitudes $A$, it is possible to map the shape of the potential and calibrate it.
This is done by analyzing the higher modes of the bead's motion that arise in response to a sinusoidal drive in a nonlinear trapping force-field.
\end{enumerate}

%%%%%%%%%%%%%%%%%%%%%%%%%%%%%%%%%%%%%%%%%%%%%%
\section{HYDRODYNAMICS}
\label{app:hydro}
Sometimes it is desirable to work with a flow-cell with \emph{open ends},
e.g.\ to facilitate the exchange of buffer solution.
When the ends are open,
the liquid between the two cover-slips generally does not move with the same amplitude
or phase as the flow-cell.

To estimate the size of this effect, we consider the motion of a liquid contained between two infinite, parallel planes which are moved identically, parallel to themselves, in a simple sinusoidal fashion.
In an experiment, two of the four sides of the flow-cell are sealed,
which increases the drag force on the liquid.
Also, the surface-tension at the openings helps to force the liquid to move with the flow-cell.
Thus, the result below for two infinite, parallel planes exaggerates the liquid's motion---it is a worst-case scenario.

The velocity of the oscillating planes (the cover-slips) is zero in the $y$ and $z$ direction, and
\be
    \vstage = \omega A \cos(\omega t)
    \enspace
\ee
in the $x$-direction, where $\omega = 2\pi\fstage$.
The only external forces on the liquid are the shear forces arising from the no-slip boundary condition
between the liquid and the accelerated planes.
The liquid moves with the accelerated planes with a time-lag determined by the balance
between the inertia of the liquid and the shear forces inside the liquid.
Thus, we see that the velocity of the liquid can be written as $\vec u(x,y,z,t) = (u(z,t),0,0)$.

The equation of motion for the liquid is obtained from the linearized Navier-Stokes equations
with constant pressure (see \cite{Landau1959}):
\be
    \frac{\partial u}{\partial t}(z,t) = \nu \frac{\partial^2 u}{\partial z^2}(z,t)
    \label{eq:u_motion}
    \enspace,
\ee
where $\nu$ is the kinematic viscosity of the liquid.
The relevant solution to this equation can be written
\bea
    u(z,t) &=& a(z) \cos(\omega t) + b(z) \sin(\omega t) \label{eq:u} \\
              &=& \Aliquid(z) \cos(\omega (t-\tliquid(z))) \label{eq:Bz}
    \enspace,
\eea
where $\Aliquid=\sqrt{a^2 + b^2}$ and $\omega \tliquid = \arctan (b/a)$.
The coefficients $a$ and $b$ can be found by substituting Eq.~(\ref{eq:u}) into Eq.~(\ref{eq:u_motion}):
\be
    a = -\frac{\nu}{\omega}  b'' \;\mbox{ and }\;
    b =  \frac{\nu}{\omega} a''  \label{eq:b1}
    \enspace,
\ee
from which we find
\be
    a'''' = -\left(\frac{\omega}{\nu}\right)^2 a
    \enspace,
    \label{eq:a4}
\ee
where $'$ indicates differentiation with respect to $z$.
The solution to Eq.~(\ref{eq:a4}) is a linear combination of
$\exp( \frac{\pm 1 \pm i}{\sqrt{2}} \sqrt{ \frac{\omega}{\nu} } z )$
whose four coefficients are determined by the symmetry requirement $a(z)=a(d-z)$
and the boundary conditions $a(0) = \omega A$ and $a''(0)= 0$:
\bea
    a(z) &=& c_1 \cos(\frac{z-d/2}{\decay}) \cosh(\frac{z-d/2}{\decay}) \nonumber \\
    &+& c_2 \sin(\frac{z-d/2}{\decay}) \sinh(\frac{z-d/2}{\decay})
\label{eq:a}
\enspace,
\eea
where
\bea
    c_1 &=& \omega A \frac{ \cos(\frac{d}{2\decay}) \cosh(\frac{d}{2\decay})}
    { \cos^2(\frac{d}{2\decay}) + \sinh^2(\frac{d}{2\decay})} \\
    c_2 &=& \omega A \frac{ \sin(\frac{d}{2\decay}) \sinh(\frac{d}{2\decay})}
    { \cos^2(\frac{d}{2\decay})  + \sinh^2(\frac{d}{2\decay})}
    \enspace ,
\eea
and
\be \label{eq:delta}
    \decay = \sqrt{2\nu / \omega}
    \enspace
\ee
 is the depth of penetration of the shear-wave into the fluid.
The expression for $b$ follows from Eq.~(\ref{eq:b1}) and is identical to Eq.~(\ref{eq:a}) except for $c_1$ and $c_2$ swapping places.
When $d \gg \decay$ the shear-wave's amplitude decreases exponentially as a function of $z$.
When $d \ll \decay$ the liquid co-moves with the planes as a solid body.

\begin{figure}[ht]
  \includegraphics[width=0.9\linewidth]{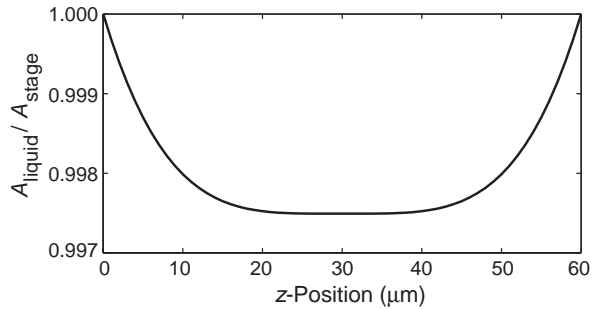}
    \caption{\label{fig:hydro_example}
    Hydrodynamic prediction of the response of the liquid to the motion of the cover-slips.
Abscissa: height above cover-slip $z$. Ordinate: Amplitude of liquid motion $\Aliquid(z)$,
 Eq.~(\ref{eq:Bz}), in units of the amplitude of the oscillating planes $A_{\rm stage}=A$.
  Parameters are the same as used in the experiment far from surfaces ($\fstage=28$\,Hz, $A=208$\,nm, $T=23.0$\degC).}
\end{figure}

With the parameters used in the experiment far from surfaces, described in Sec.~\ref{sec:test}, i.e., $\fstage = 28$\;Hz,
the penetration depth $\decay = 103$\;$\mu$m is larger than the thickness of the flow-cell $d\simeq 60$\;$\mu$m.
With these parameters, the amplitude of the liquid's motion differ at most 0.25\% from the amplitude of  the stage's motion, see Fig.~\ref{fig:hydro_example}.
This result is independent of the amplitude $A$ but is quite sensitive to $d$ and $\fstage$.
Thus, if working away from the surface it is important to choose $\fstage$ so that it matches the thickness of the cell,
 i.e., so that $d/\decay < 1$.
%----------------------------------------------------------------------
%------------------------------------------------------------------------

\begin{figure}[ht]
   \includegraphics[width=0.9\linewidth]{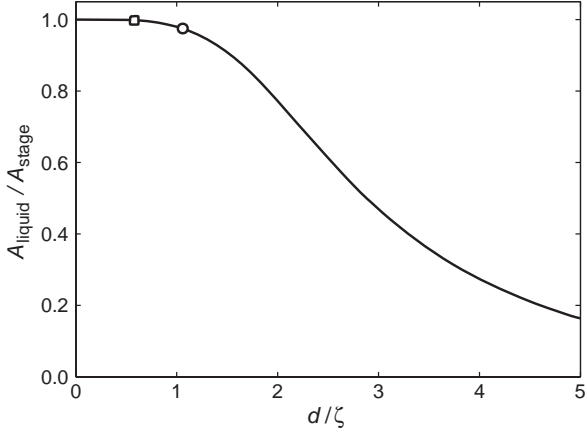}
    \caption{\label{fig:hydro_d_delta}
  Hydrodynamic prediction of the amplitude of liquid motion midway between the two cover-slips.
    Abscissa: The distance between two planes $d$, measured in units of the penetration depth $\decay$, Eq.~(\ref{eq:delta}).
    Ordinate: Amplitude of the liquid-oscillations midway between the two planes $\Aliquid(\frac{d}{2})$ Eq.~(\ref{eq:Bhalf}), in units of the amplitude of the oscillating planes $A_{\rm stage} = A$.
    Notice the plateau at unity for small values of $d/\decay$
    where the fluid moves with the planes,
    and the exponential decay to zero at larger values where the liquid no longer follows the planes.
    The values of $d/\decay$ for the experiments in bulk and close to the surface are marked by a square and a circle, respectively.}
\end{figure}

It is illustrative to calculate the velocity of the liquid midway between the two planes,
since this is where the motion of the liquid differs the most from that of the  planes:
\bea
   u(\frac{d}{2},t) &=& c_1 \cos(\omega t) + c_2 \sin(\omega t) \\
                 &=& \Aliquid(\frac{d}{2}) \cos(\omega (t-\tliquid(d/2))) \label{eq:Bhalf}
   \enspace,
\eea
where now $\Aliquid(\frac{d}{2}) = \sqrt{c_1^2 + c_2^2}$ and $\omega \tliquid(d/2) = \arctan(c_2/c_1)$.
When the cover-slips are close together or the drive frequency is very low $\frac{d}{2\decay} \ll 1$,
and we have $u(\frac{d}{2},t) = \vstage(t)$, i.e., the fluid co-moves with the planes.
If the cover-slips are far apart or the drive-frequency is high  Eq.~(\ref{eq:Bhalf}) becomes:
\be
    u(\frac{d}{2},t) = 2\omega A  e^{-\frac{d}{2\decay}} \cos(\omega t - d/2\decay)
    \mbox{, } d \gg 2\decay
    \enspace,
\ee
i.e., the amplitude decreases exponentially as  $d/\decay$ grows and the motion of the liquid is phase-shifted relative to the motion of the planes.
These behaviors are illustrated in Fig.~\ref{fig:hydro_d_delta}.
From this we see the importance of choosing a drive-frequency small enough for the fluid to follow the flow-cell in the region where the trap is to be calibrated.
%----------------------------------------------------------------------
%------------------------------------------------------------------------

%%%%%%%%%%%%%%%%%%%%%%%%%%%%%%%%%%%%%%%%%%%%%%%
%%%%%%%%%%%%%%%%%%%%%%%%%%%%%%%%%%%%%%%%%%%%%
\section{Power spectral density for finite measurement time}
\label{app:psd_rect}
In Section~\ref{sec:theory} we calculated the PSD for the beads' motion,
assuming infinite measurement time as well as continuous sampling.
In a real experiment, data are collected for a finite time $\Tmsr$
and with a finite sampling frequency $\fsample$.
In what follows we are going to assume continuous sampling in time,
since the \emph{aliasing} caused by finite sampling time (i) is irrelevant for the discussion here,
(ii) is easily accounted for~\cite{Berg-Sorensen2004},
and (iii) does not occur at all,
if over-sampling data acquisition electronics is used.
Readers feeling uneasy about the idea of continuous sampling
need only remind themselves that it is a mathematical ideal,
a limit that is achieved in practice by choosing  $\fNyq =\fsample/2  \gg \fc$
and, if aliasing occurs, by leaving frequencies with significant aliasing out of the analysis.
Equations~(H5) and (H6) in \cite{Berg-Sorensen2004}
indicate which frequencies have significant aliasing
for a given desired accuracy.

For the finite-time continuous-time Fourier transform of Eq.~(\ref{eq:bead_pos})
we now have
\be
    \tilde{x}_k = \int_{-\Tmsr/2}^{\Tmsr/2}  \mbox{d}t \, e^{i 2 \pi f_k t} x(t)
    \enspace ,
    \label{eq:x_fourier_rect}
\ee
and the expectation value for the one-sided ($f_k \ge 0$) PSD becomes:
\be
    \label{eq:rect_power}
    P_k(\Tmsr) = \frac{ 2\langle | \tilde{x}_k |^2 \rangle }{ \Tmsr }
    = \frac{ D + (\fstage r_k A)^2 \Tmsr}{\pi^2(f_k^2 + \fc^2)}  \
    \enspace ,
\ee
where
\be
    r_k = \frac{1}{\sqrt 2} \frac{ \sin(\pi(f_k-\fstage)\Tmsr) }{(f_k-\fstage)\Tmsr }
    \label{eq:rect_peak}
    \enspace,
\ee
gives the shape of the spike in the PSD,
referred to as \emph{leakage} \cite{Harris1978}; see \cite[Fig.~5]{Berg-Sorensen2003} for peak shapes.
Here, $f_k = k/\Tmsr$ where $k$ is an integer.
We have ignored cross terms of the type
$\sin(\pi(f_k-\fstage)\Tmsr)\sin(\pi(f_k+\fstage)\Tmsr)$ in Eq.~(\ref{eq:rect_peak}),
as they are typically several orders of magnitude smaller than the terms retained.

The recording time $\Tmsr$ is easily chosen to be an integer multiple of the period of the stage's motion $\fstage$.
It can be done even after recording was finished,
simply by discarding an incomplete stage period from the recording.
A simplification is thereby achieved,
$\fstage = f_k$ for some integer $k$,
so the spike consists of a single datum
\be
    r_k = \frac{ \pi }{ \sqrt{2} } \, \delta_{f_k,\fstage}
\ee
and the PSD becomes
\be
    P_k(\Tmsr) = \frac{ D }{ \pi^2(f_k^2 + \fc^2) } +
    \frac{ \frac{1}{2}\fstage^2 A^2}{f_k^2 + \fc^2} \Tmsr \, \delta_{f_k,\fstage}
    \enspace
    \label{eq:finite_psd}
\ee
where $\delta_{f_k,\fstage}$ is Kronecker's delta.
This is the discrete version of the expression given in Eq.~(\ref{eq:inf_power}).
We also see that the height of the spike depends linearly on the measurement time $\Tmsr$.
When the measurement time is increased the spike approaches Dirac's delta function
\be
    \Tmsr \, \delta_{f_k,\fstage} \rightarrow \delta(f - \fstage),\,\, \Tmsr \rightarrow \infty
    \enspace
\ee
and the expression in Eq.~(\ref{eq:inf_power}) is regained.

%%%%%%%%%%%%%%%%%%%%%%%%%%%%%%%%%%%%%%%%%%%%%%%%%%%%%%%%%%%%%%%%%%%%%%%%%%%%%%%%%%%%%%
\section{Hydrodynamically correct Power spectrum analysis}  \label{app:Phydro}

No general statement can be made regarding which accuracy to expect from a Lorentzian fit.
The importance of omitted effects depends on several circumstances~\cite{Berg-Sorensen2004,Berg-Sorensen2006_preprint}.
As a rule of thumb, we recommend
the use of the hydrodynamically correct theory for Brownian motion,
if less than 10\% systematic error is desired.
User-friendly software exists that calibrates with the full theory~\cite{Tolic-Norrelykke2004,Hansen2006},
and one is left with fewer concerns regarding reliability of results.

%%%%%%%%%%%%%%%%%%%%%%%%%%%%%
\subsection{Hydrodynamically correct power spectrum}
The hydrodynamically correct power-spectrum of Brownian motion
of a micro-sphere that is trapped by a Hookean force,
is
\be \label{eq:Phydro}
  \Phydro(f,R/\ell) = \Pthermhydro + \Pdrivenhydro
  \enspace,
\ee
where
\bea \label{eq:Pthermhydro}
   \lefteqn{ \Pthermhydro(f,R/\ell) = }&&  \\
   &&\frac{ (\Do/\pi^2) \, \real \{ \gfl/\go \} }
    { (\fco + f\,\imag\{ \gfl/\go \} - f^2/\fmo)^2  +  (f\,\real\{ \gfl/\go \})^2 }
    \enspace. \nonumber
\eea
and
\bea \label{eq:Pdrivenhydro}
   \lefteqn{ \Pdrivenhydro(f,R/\ell) = }&&  \\
   &&\frac{ \frac{1}{2} (A \fstage | \gfl/\go |)^2 \, \delta(f-\fstage)}
    { (\fco + f\,\imag\{ \gfl/\go \} - f^2/\fmo)^2  +  (f\,\real\{ \gfl/\go \})^2 }
    \enspace. \nonumber
\eea
Equation~(\ref{eq:Pthermhydro}) is the same expression as given in \cite[Eq.~34]{Berg-Sorensen2004}
(Note however \cite{footnote3}).
Here  $\gfl(f,R/\ell)$ is the frequency-specific drag coefficient that results
from solving Stokes equations for the micro-sphere undergoing linear harmonic motion with frequency $f$.
We assume this motion takes place parallel to an infinite planar surface
that is located a distance $\ell$ from the center of the microsphere with radius $R$,
and give an approximate expression for $\gfl(f,R/\ell)$ below.
Using the notation $\go = \gfl(0,0)=6\pi \rho \nu R$,
where $\rho$ is the density of the surrounding liquid and $\nu$ its kinematic viscosity,
the parameters in Eqs.~(\ref{eq:Pthermhydro},\ref{eq:Pdrivenhydro}) are
$\Do=\kB T/\go$,
$\fco = \kappa /( 2 \pi \go )$, and
$\fmo = \go/( 2 \pi m )$ with $m$ the mass of the micro-sphere.
All three parameters are independent of $\ell$.
They describe bulk values, while all $\ell$-dependence of the PSD is found in $\gfl(f,R/\ell)/\go$.
In this manner,  consistency of experimental power spectra
that are recorded with the same bead and trap
at different distances to a surface, is very easy to check:
The values returned for $\Do$ and $\fmo$ should all be the same,
while variation in $\fco$ arise from variation due to optical aberrations (that depend on $\ell$~\cite{Schaeffer_in_prep}).

%%%%%%%%%%%%%%%%%
\subsection{The drag coefficient}
The drag coefficient $\gfl(f,R/\ell)$ is not known exactly for $R/\ell>0$.
Deep in bulk, $R/\ell=0$,
and one has Stokes' exact result \cite{Stokes1851}
\bea \label{eq:stokes}
\lefteqn{\gfl(f,0) = }&&\\
&&\gstokes(f) \equiv \go \left( 1 + (1-i) \sqrt{\frac{f}{f_\nu}} - i \frac{2}{9}\frac{f}{f_\nu} \right)
\nonumber
\enspace,
\eea
where $f_\nu = \nu/(\pi R^2)$.
At zero frequency one has Fax\'en's approximate result \cite{HappelBrenner,Faxen1921},
\bea \label{eq:faxen}
\lefteqn{\gfl(0,R/\ell) = }&&\\
&&\gfaxen(R/\ell) \equiv
\frac{\go}{1-\frac{9R}{16\ell}+\frac{R^3}{8\ell^3}-\frac{45R^4}{256\ell^4}-\frac{R^5}{16\ell^5}+\ldots}
\enspace.  \nonumber \eea
An exact result exists \cite{Oneill1964} and has been shown numerically \cite{Goldman1967} to agree with Eq.~(\ref{eq:faxen}) to within 1\% for $\ell/R \gtsim 1.4$.
 
An approximate expression for  $\gfl(f,R/\ell)$ is given in  \cite[Eq.~(33)]{Berg-Sorensen2004}.
At zero frequency, it reproduces Fax\'en's result up to and including its first order term in $R/\ell$.
In the spirit of Pad\'e approximants,
we rearranged the terms of  \cite[Eq.~(33)]{Berg-Sorensen2004}
to
\bea \label{eq:gammapade}
\lefteqn{ \gfl(f,R/\ell) =}&&  \\
   && \frac{ \gstokes(f) }
    { 1 - \frac{9}{16}\frac{R}{\ell}
    \left[ 1 - \frac{1-i}{3} \sqrt{\frac{f}{f_\nu}}   + \frac{2if}{9f_\nu} -\frac{4}{3}
     (1 - e^{ -(1-i)\frac{2\ell-R}{\delta}} ) \right]+\ldots} \nonumber
\eea
where $\delta = R\sqrt{f_\nu/f}$,
and the series in $R/\ell$ in the denominator is not known beyond the explicitly shown terms.
For $f=0$ this expression reproduces Fax\'en's result up to and including its \emph{second} order term in $R/\ell$:
For $\ell/R > 1.5$ it deviates by less than 0.5\% from Fax\'en's
result, and for $\ell/R > 5$ the deviation is smaller than 0.1\%.
This is a significantly better approximation to Fax\'en's result, so
it must also be a significant improvement at low frequencies, and
possibly at all frequencies.

We found that Eq.~(\ref{eq:Pthermhydro}) agrees with our experimental PSDs
down to significantly shorter distances from the surface
when $\gfl(f,R/\ell)$ in Eq.~(\ref{eq:gammapade}) is used
instead of  Eq.~(33) in~\cite{Berg-Sorensen2004}, 
see \cite{footnote4,footnote5}.
%%%%%%%%%%%%%%%%%%%%%%%%%%%%%%%%
\subsection{How to fit the $\ell$-dependent power spectrum}

In order to fit the theoretical power spectrum in Eqs.~(\ref{eq:Phydro}--\ref{eq:Pdrivenhydro})
to the experimental one, the distance $\ell$ to the surface must be known.
We do know the \emph{differences} between the various distances
at which we recorded power spectra.
It is only the location of the coverslip surface that is unknown.

This location can be determined in several ways:
From the fluorescence induced by an evanescent wave,
from the effect of the hydrodynamic interaction of the trapped bead with the surface,
or by analyzing interference or diffractive patterns~\cite[references therein, and \cite{Hansen2005,Neuman2005}]{Neuman2004}.
Yet another method relies on pulling on DNA constructs along the axial direction \cite{Deufel2006}.
Most of these methods require a modification of the experimental setup.
We used the hydrodynamic interaction of the bead with the surface,
similar to that described in \cite{Wang1997}
but taking advantage of an improved understanding of the focal shift,
as detailed in \cite{Schaeffer_in_prep}.
This method requires no extra instrumentation.

In principle, one might fit  Eqs.~(\ref{eq:Phydro}--\ref{eq:Pdrivenhydro}) simultaneously
to all power spectra recorded with a given bead in the trap,
and let the location of the surface be a fitting parameter.
We used a simpler procedure, based on the observation that
the distance to the surface is needed only with moderate precision in order to
determine the fitting parameters  $\Do$ and $\fco$ with high precision.
This is because the distance occurs only in Eq.~(\ref{eq:gammapade}) in terms
that play a subdominant role in Eqs.~(\ref{eq:Phydro}--\ref{eq:Pdrivenhydro}).

We fitted Eq.~(\ref{eq:Pthermhydro}) to the thermal background and compared the results from inserting either $\gamma = \gfaxen$ (turning Eq.~(\ref{eq:Pthermhydro}) into a Lorentzian) or $\gamma = \gfaxen \times \gstokes / \go$.
The distance-dependent results for $\fc$ and $D$ were found to agree within 1\% for the two approaches.
This agreement is partly artificial:
We described the parasitic filtering in our diode with \cite[Eq.~(35)]{Berg-Sorensen2004},
fitting its two parameters.
This characteristic function absorbed most of the difference between
$\go$ and $\gstokes(f)$.

Einstein's relation then gives a distance-dependent
$\gac=\kB T/D$, while Fax\'en's law should give the distance dependence,
$\gac \approx \gfaxen(R/\ell)$.
We therefore fitted Fax\'en's law to our $\ell$-dependent values for $\gac$,
using the location of the surface as a fitting parameter.
The good fits confirmed the procedure~\cite{Schaeffer_in_prep}.

With $\ell$ thus known, we know $\gfl(f,R/\ell)$ in Eq.~(\ref{eq:gammapade}),
and can fit the experimental PSDs with the more correct theoretical PSD
that results from using Eq.~(\ref{eq:gammapade}) in Eq.~(\ref{eq:Pthermhydro}),
and  $\Do$ and $\fco$ as fitting parameters  with their \emph{original} interpretation:
$\Do=\kB T/\go$ and $\fco=\kappa/(2\pi \go)$.

Note that although $\fnu$ depends on the radius of the bead
and on the kinematic viscosity of the surrounding fluid,
it is systematically correct to insert the manufacturer's value for the radius
and the room temperature value for the viscosity in $\fnu$.
This is because in Eq.~(\ref{eq:gammapade}) these small errors occur
in a term that is already a small correction.

%%%%%%%%%%%%%%%%%%%%%%%%%%%%%
\subsection{Hydrodynamically correct calibration factor}
The hydrodynamically correct theory described above gives
\bea \label{eq:Wthhydro}
\lefteqn{ W_{\rm th, hydro} = \int \Pdrivenhydro \mathrm{d}f = } && \\
&& \frac{ \frac{1}{2} A^2 |\gfl/\go|^2}
    { (\fco/\fstage + \imag\{ \gfl/\go \} - \fstage/\fmo)^2  +  (\real\{ \gfl/\go \})^2 }
    \nonumber
    \enspace.
\eea
This expression does not change the calibration factor $\beta$ by much,
compared to what Eq.~(\ref{eq:Wth}) would give.
This is because Einstein's approximate theory for Brownian motion
that lead to Eq.~(\ref{eq:Wth}),
is a low-frequency approximation,
and $\fstage$ is a low frequency compared to the characteristic frequency
$\fnu$ of the frequency dependent drag.
At room temperature
$(\fstage/\fnu)^{1/2}=7.2\cdot 10^{-3}$ for the 1.54\,$\mu$m diameter beads we drove at $\fstage=28$\,Hz,
and $(\fstage/\fnu)^{1/2}=2.6\cdot 10^{-3}$ for the 528\,nm diameter beads we drove at $\fstage=32$\,Hz.
Inertial effects are two orders of magnitude smaller~\cite{Berg-Sorensen2004}.
Comparing with Eq.~(\ref{eq:Wth}),
we see that, in bulk, the hydrodynamically correct theory in our case introduces corrections of 1.5\% and 0.5\%, respectively, in the interpretation of the power in the spike.
In contrast, near a surface the corrections to Eq.~(\ref{eq:Wth}) are of the opposite sign and a smaller magnitude than the corrections in bulk: For $\ell/R = 10$, $\fstage=32$\,Hz, $\fco=2000$\,Hz, and $2R=528$\,nm the correction is $-0.25\%$.

%%%%%%%%%%%%%%%%%%%%%%%%%%%%%%%%%%%%
%\clearpage
% \section*{References}
%\bibliographystyle{prsty}
%\bibliography{/Users/simon/00Simon/Project_AC_Calibration/bibliography_ac_calibration_bibdesk}% TATA.bib

\begin{thebibliography}{10}
\bibitem{footnote3}
Equation~(34) in \cite{Berg-Sorensen2004} contains a typographical error: The
  parameter $\fm$ in this equation is $\fmo=\go/(2\pi m)$, i.e., not the same
  as the parameter $\fm$, which occurs in Eq.~(32) in \cite{Berg-Sorensen2004}.
  Systematic naming is achieved, if the latter is referred to as
  $f_{m^*,0}=\go/(2\pi m^*)$, i.e., only its name is changed, not its
  definition in terms of $m^*= m+2\pi \rho R^3/3$, where $\rho$ is the mass
  density of the liquid \cite{Berg-Sorensen_toappear}. This error is \emph{not}
  repeated in the programs described in \cite{Tolic-Norrelykke2004,Hansen2006}.
\end{thebibliography}

\end{document}